\documentclass[
letter,
showpacs,
floatfix,
aps,
prl,
amsmath,
twocolumn,
superscriptaddress,
tightenlines
groupaddress,
]{revtex4}

\usepackage{graphicx}
\usepackage{dcolumn}
\usepackage{amsmath}
\usepackage{amsfonts}
\usepackage{bm}
\usepackage{times}

\def\eac{\epsilon_{\mbox{{\scriptsize ac}}}}
\def\edc{\epsilon_{\mbox{\scriptsize dc}}}

\def\pac{\phi_{\mbox{\scriptsize ac}}}

\def\oc{\omega_{\mbox{\scriptsize {C}}}}

\def\rc{R_{\mbox{\scriptsize {C}}}}

\bibpunct{[}{]}{,\!}{n}{,}{,} 
\begin{document}

\title{\fontfamily{ptm}\selectfont
Microwave Photoresistance in dc-driven 2D Systems at Cyclotron Resonance Subharmonics
}
\author{A.\,T. Hatke}
\affiliation{School of Physics and Astronomy, University of Minnesota, Minneapolis, Minnesota 55455, USA} 
\author{H.-S. Chiang}
\affiliation{School of Physics and Astronomy, University of Minnesota, Minneapolis, Minnesota 55455, USA} 
\author{M.\,A. Zudov}
\email[Corresponding author: ]{zudov@physics.umn.edu}
\affiliation{School of Physics and Astronomy, University of Minnesota, Minneapolis, Minnesota 55455, USA} 
\author{L.\,N. Pfeiffer}
\affiliation{Bell Labs, Alcatel-Lucent, Murray Hill, New Jersey 07974, USA}
\author{K.\,W. West}
\affiliation{Bell Labs, Alcatel-Lucent, Murray Hill, New Jersey 07974, USA}

\begin{abstract}
We study microwave photoresistivity oscillations in a high mobility two-dimensional electron system subject to strong dc electric fields.
We find that near the second {\em subharmonic} of the cyclotron resonance the frequency of the resistivity oscillations with dc electric field is twice the frequency of the oscillations at the cyclotron resonance, its harmonics, or in the absence of microwave radiation.
This observation is discussed in terms of the microwave-induced sidebands in the density of states and the interplay between different scattering processes in the separated Landau level regime. 
\end{abstract} 
\pacs{73.40.-c, 73.21.-b, 73.43.-f}
\maketitle

Over the recent years many remarkable phenomena, other than conventional Shubnikov-de Haas oscillations \citep{shubnikov:1930}, have been discovered in high Landau levels of two-dimensional electron systems (2DES).
A lot of attention has been paid to microwave-induced resistance oscillations (MIRO)\,\citep{miro:exp} and zero-resistance states\,\citep{zrs} formed at the MIRO minima.
Another notable effect is Hall field-induced resistance oscillations (HIRO) emerging under strong dc electric field\,\citep{yang:2002a,zhang:2007a}.
Recently, it was demonstrated that dc field also induces strong resistance suppression\,\citep{zhang:2007a,zhang:2007b} and zero-differential resistance states\,\citep{bykov:2007,zhang:2008}.

Stepping from the inter-Landau level transitions due to microwave absorption and/or impurity scattering, MIRO and HIRO are controlled by simple parameters, $\eac\equiv\omega/\oc$ and $\edc\equiv eE(2\rc)/\hbar\oc$, respectively.
Here, $\omega=2\pi f$ is the microwave frequency, $\oc=eB/m^*$ is the cyclotron frequency, $E$ is the Hall electric field, and $2\rc$ is the cyclotron diameter.
MIRO maxima$^+$ and minima$^-$ are found at $\eac^{\pm}\simeq n \mp \pac,\,\pac \leq 1/4$ ($n \in \mathbb{Z}^+$)\,\citep{zudov:2004,studenikin}, while
HIRO maxima (minima) occur near integer (half-integer) values of $\edc$ \cite{hiro:disp}.
To explain MIRO, ``displacement''\,\citep{miro:disp} and ``inelastic''\,\citep{miro:in} models have been proposed. 
Until recently it was believed that the ``inelastic'' mechanism, originating from the oscillatory correction to the electron distribution, dominates over the ``displacement'' mechanism, which is based on microwave-assisted scattering off of impurities.
However, recent experiments \citep{hatke:2009} have shown that the ``displacement'' mechanism cannot be ignored.

Remarkably, in addition to MIRO at integer $\eac$, many experiments \citep{zudov:2004,fmiro:exp} reported similar features near certain {\em fractional} values of $\eac =$ 1/2, 3/2, 5/2, 1/3, 2/3.
These oscillations were initially explained by a multiphoton ``displacement'' mechanism \citep{lei:2006a} which relies on a single electron simultaneously absorbing multiple photons.
However, recently \citep{dmitriev:2007b} it was argued that in the regime of separated Landau levels ``inelastic'' multiphoton mechanism dominates the response while at the crossover from separated to overlapped Landau levels, two single-photon ``inelastic'' mechanisms become important.
The first of these mechanisms is based on a resonant series of consecutive single-photon transitions \citep{pechenezhskii:2007}.
The other originates from microwave-induced sidebands in the density of states of the disorder-broadened Landau levels \citep{dmitriev:2007b}.

When a 2DES is subject to microwaves and a strong dc electric field simultaneously \citep{miro:dc,miro:dc:th}, the resistivity oscillations are governed by remarkably simple combinations of ac and dc parameters, namely $(\eac\pm\edc)$.
More specifically, the maxima in the differential resistivity are well described by $\eac+\edc \simeq n$ and 
$\eac-\edc\simeq m-1/2\,(m \in \mathbb{Z}^+)$ \citep{miro:dc}. 
These two conditions indicate that not only scattering in the direction {\em parallel} to the dc field needs to be maximized (first condition), but also that scattering {\em antiparallel} to the dc field, has to be minimized (second condition).
Therefore, in the presence of microwaves and strong dc electric field the resistivity is explained by the ``displacement'' mechanism, {\em i.\,e.} in terms of combined inter-Landau level transitions which are composed of both microwave absorption and impurity scattering.
Somewhat similar effect of dc electric field on resonant acoustic phonon scattering \citep{zudov:2001b} was recently observed \citep{zhang:2008}. 

In this Letter, we study the effect of dc electric field on microwave photoresistivity near cyclotron resonance subharmonics of a high mobility 2DES.
Surprisingly, the nonlinear response around the second cyclotron resonance subharmonic, $\eac = 1/2$, closely mimics the response near cyclotron resonance, $\eac = 1$, despite a factor of two difference in magnetic fields.
In other words, differential resistivity at $\eac\simeq 1/2$ oscillates with {\em double} frequency, exhibiting maxima not only at integer but also at half-integer values of $\edc$.
This is in vast contrast with the HIRO-like $\cos(2\pi\edc)$ dependence observed at {\em all} cyclotron resonance harmonics, $\eac=n$ \citep{miro:dc}.
We discuss this observation in terms of  recently proposed \citep{dmitriev:2007b} microwave-induced sidebands in the density of states and in terms of the competition between different scattering processes in the separated Landau level regime.
It is important to extend existing theories \citep{lei:2006a,dmitriev:2007b,pechenezhskii:2007} into the non-linear regime which, when combined with our results, will help to identify the underlying microscopic mechanisms for fractional oscillations and frequency doubling.

Our experiment was performed on a 100 $\mu$m-wide Hall bar etched from a symmetrically doped GaAs/AlGaAs quantum well.
After brief low-temperature illumination with visible light, density and mobility were $n_e\simeq 3.8\times 10^{11}$ cm$^{-2}$ and $\mu\simeq 1.3\times 10^{7}$ cm$^{2}$/Vs, respectively.
All the data were recorded under continuous irradiation by $f=27$ GHz microwaves in a Faraday geometry at $T\simeq$ 1.5 K.
Differential resistivity, $r_{xx}\equiv dV/dI$, was measured using a quasi-dc (a few Hertz) lock-in technique.

\begin{figure}[t]
\includegraphics{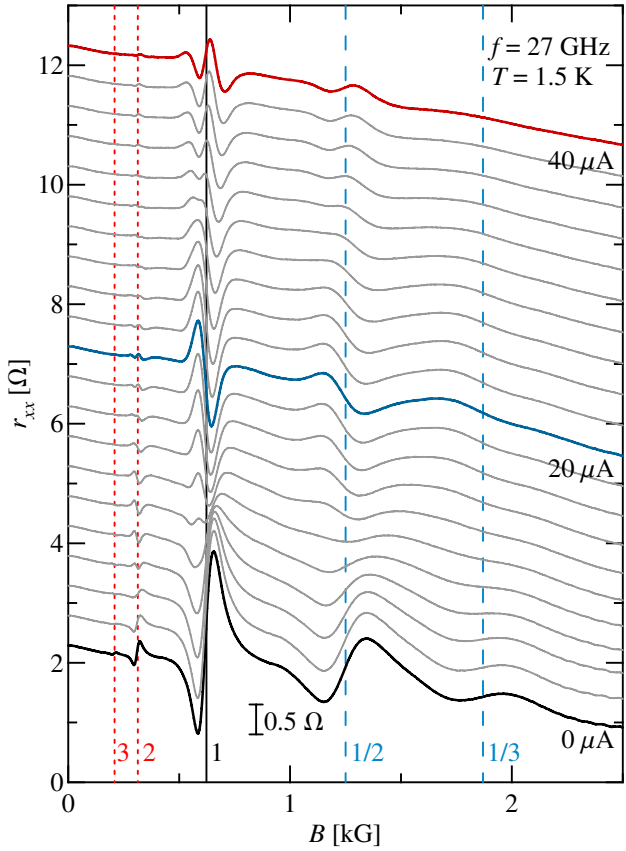}
\caption{(color online)
Differential magnetoresistivity $r_{xx}(B)$ under microwave illumination of $f= 27$ GHz measured at $T=1.5$ K for $I$ from 0 to 40 $\mu$A in steps of 2 $\mu$A.
The traces are vertically offset by $0.5\,\Omega$.
Vertical dotted (dashed) lines are drawn at $\eac=2,3$ ($\eac=1/2,1/3$) marking harmonics (subharmonics) of the cyclotron resonance (solid line).
}
\vspace{-0.15 in}
\label{fig1}
\end{figure}
In Fig.\,\ref{fig1}\, we present differential magnetoresistivity $r_{xx}(B)$ at fixed current $I$ ranging from 0 to 40 $\mu$A in steps of 2 $\mu$A.
The traces are vertically offset for clarity by $0.5\,\Omega$.
At zero dc bias (bottom curve) the data show MIRO appearing as maximum-minimum pairs positioned symmetrically about cyclotron resonance harmonics, $\eac=2$ and $\eac=3$ (cf.,\,dotted lines), and subharmonics, $\eac=1/2$ and $\eac=1/3$ (cf.,\,dashed lines).
While the oscillation near the cyclotron resonance, $\eac=1$, is the strongest, the MIRO amplitude near $\eac=1/n$ greatly exceeds that at $\eac=n$, for $n=2,\,3$. 
\begin{figure}[t]
\includegraphics{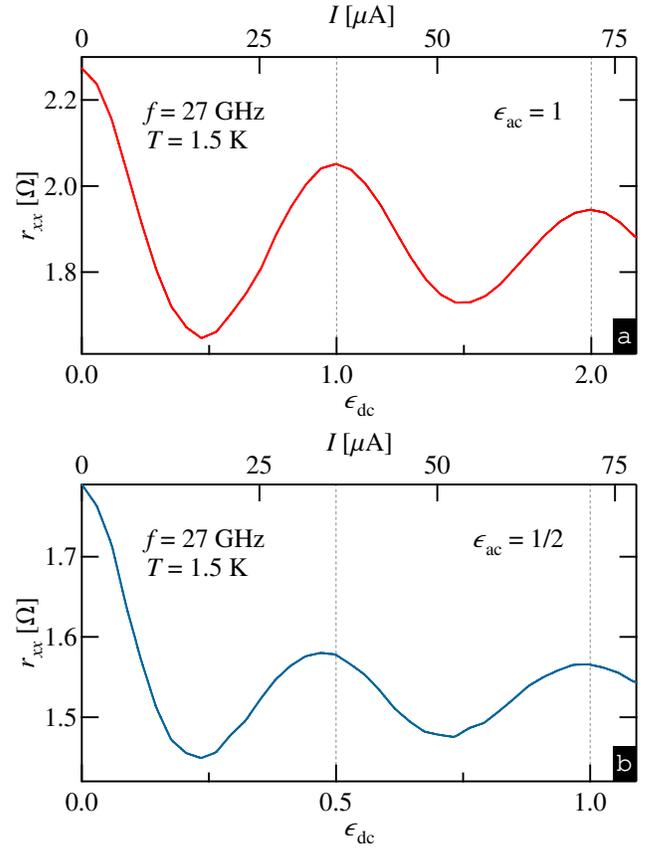}
\caption{(color online)
(a)[(b)] Differential resistivity $r_{xx}$ vs $\edc$ (bottom axis) and $I$ (top axis) at $\eac\simeq 1$ [$\eac\simeq 1/2$]. 
}
\vspace{-0.15 in}
\label{fig2}
\end{figure}

We now compare the effect of the dc current on the oscillation near $\eac=1/2$ to the oscillation near $\eac=1$.
From previous studies \citep{miro:dc} we expect that maxima (minima) will continuously evolve into the minima (maxima) and back into the maxima (minima) periodically with increasing current.
Indeed, extrema near $\eac=1$ switch places at $I\simeq 20$ $\mu$A and, at a higher current, $I\simeq 40$ $\mu$A, switch back to the zero-bias situation.
Unexpectedly, maximum and minimum near $\eac=1/2$ show very similar behavior in the {\em same} range of currents;
the maximum (minimum) becomes a minimum (maximum) at 20 $\mu$A and then reverts back to a maximum (minimum) at 40 $\mu$A.
This is quite remarkable, since naively one would expect that similar evolution at $\eac=1/2$ would require twice as high $I$ compared to $\eac=1$ because $\edc\propto I/B$ and the magnetic field is twice as high.
We recall that in all previous studies the differential resistivity followed $\cos[2\pi(\edc+\varphi)]$, where  $\varphi$ could assume any value depending on $\eac$.
In other words, microwaves were known to affect only the phase of the dc-induced oscillations and the period stayed the same as without microwaves.

To confirm our observation, we present in Fig.\,\ref{fig2} \,(a) and (b) $r_{xx}$ as a function of $I$ (top axis) and $\edc$ (bottom axis) at fixed $\eac\simeq 1$ and $\eac\simeq 1/2$, respectively.
The data for $\eac=1$ show strong HIRO-like periodic oscillations following the expected $\cos(2\pi\edc)$ dependence, as observed in previous experiments \citep{miro:dc}.
Similar data, obtained at the second subharmonic, $\eac=1/2$, also show well developed, periodic oscillations with the {\em same} period in terms of $I$ despite twice as high the magnetic field.
After conversion to $\edc$ the frequency of these oscillations is doubled compared to the $\eac=1$ case and the peaks are found near both $\edc=1/2$ and $\edc=1$.

We further notice that the overall shapes of the waveforms presented in Fig.\,\ref{fig2} \,(a) and (b) are very similar and that the peak at $\edc=1/2$ does not stand out in any significant way, being comparable in magnitude to the fundamental $\edc=1$ peak.
Since without microwaves one would observe HIRO-like oscillations and the peak near $\eac=1/2$ would disappear we conclude that this peak originates from the ``combined'' action by microwaves and dc electric field.

\begin{figure}[t]
\includegraphics{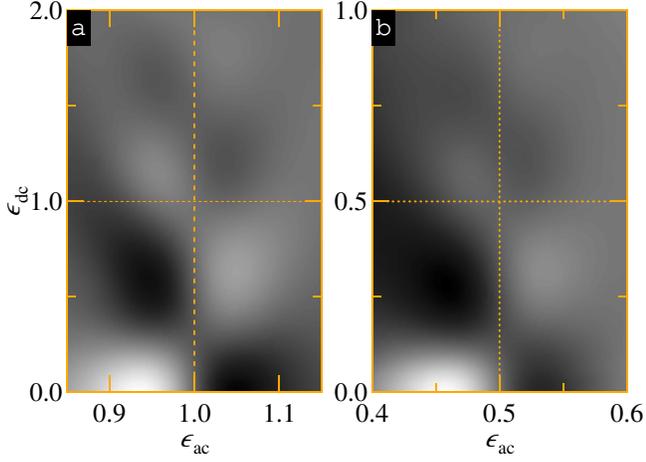}
\caption{(color online)
(a)[(b)] Grey-scale plot of $r_{xx}$ in the $(\eac,\edc)$-plane, near $(\eac,\edc)=(1,1)$ [$(\eac,\edc)=(1/2,1/2)$]. Light (dark) corresponds to high (low) $r_{xx}$ values.
}
\vspace{-0.15 in}
\label{fig3}
\end{figure}
To compare the behavior in the finite range of parameters in the vicinity of $(\eac,\edc)=(1,1)$ and $(\eac,\edc)=(1/2,1/2)$, we construct differential resistivity maps and present the results in Fig.\,\ref{fig3}\,(a) and Fig.\,\ref{fig3}\,(b).
Careful examination of Fig.\,\ref{fig3} reveals a remarkable degree of similarity between these maps, apart from different axes scaling.
Indeed, the patterns formed by the differential resistivity in these maps closely mimic each other showing a topologically identical landscape.
In particular, both Fig.\,\ref{fig3} (a) and (b) reveal three saddle points near $(\eac,\edc)=(1,1/2),(1,1),(1,3/2)$ and $(\eac,\edc)=(1/2,1/4),(1/2,1/2),(1/2,3/4)$, respectively.
Observed similarity suggests that conditions for the $r_{xx}$ maxima$^+$ and minima$^-$ near $\eac=1/2$ can be obtained by simple rescaling of the conditions for integer MIRO, $(\eac,\edc)^+\simeq(n\pm 1/4,m\mp 1/4)$ and $(\eac,\edc)^-\simeq(n\pm 1/4,m\pm 1/4)$ \citep{miro:dc}.
Thus, in the vicinity of the second cyclotron resonance subharmonic, $\eac=1/2$, positions of the differential resistivity maxima$^+$ and minima$^-$ are given by:
\begin{align}
(\eac,\edc)^+\simeq(1/2\pm 1/8,1/2\mp 1/8),\notag\\
(\eac,\edc)^-\simeq(1/2\pm 1/8,1/2\pm 1/8).
\label{res2}
\end{align}
We emphasize that Eq.\,(\ref{res2}) is expected to hold only in the regime of the overlapped Landau levels and low microwave intensities, which, strictly speaking, is not satisfied in our experiment.
Near $\eac=1/2$ we estimate $\oc\tau_q \simeq 6$ ($\tau_q$ is the quantum lifetime) and observe a significantly reduced MIRO phase $\pac \simeq 0.036 \ll 1/8$, which should modify the ``1/8-shifts'' in Eq.\,(\ref{res2}).
However using Eq.\,(\ref{res2}) simplifies the discussion and it should still correctly describe the saddle points and the relative positions of the extrema. 

\begin{figure}[t]
\includegraphics{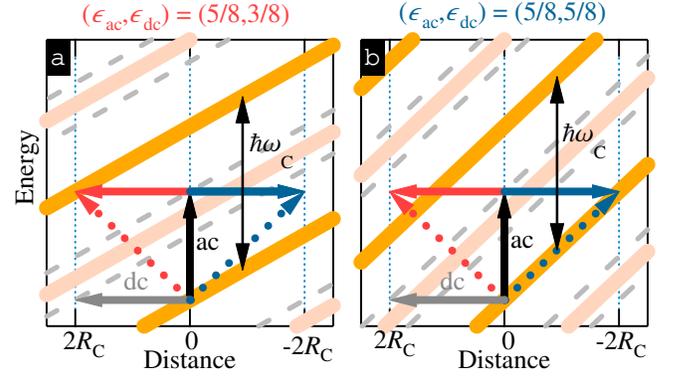}
\caption{(color online)
Combined transitions (dotted arrows) between Landau levels (dark lines) and theoretically predicted microwave-induced sidebands \citep{dmitriev:2007b} (light lines) consist of microwave absorption (vertical arrows) and impurity scattering (horizontal arrows) for (a) a maximum at $(\eac,\edc)=(5/8,3/8)$ and (b) a minimum at $(\eac,\edc)=(5/8,5/8)$.
}
\vspace{-0.15 in}
\label{fig4}
\end{figure}
Observed frequency doubling can be understood in terms of microwave-induced sidebands in the density of states which form around $\varepsilon_N \pm \hbar \omega$, where $\varepsilon_N=\hbar\oc(N+1/2)$ is the center of the $N$-th Landau level \citep{dmitriev:2007b}.
At $\eac \simeq 1/2$, sidebands from higher and lower Landau levels strongly overlap and produce a maximum in the density of states exactly in the center of the cyclotron gap.
As a result, the period of the density of states is reduced by a factor of two and the situation becomes similar to that near the cyclotron resonance, $\eac \simeq 1$.
This is exactly what we have observed in our experiment.

The situation is illustrated in Fig.\,\ref{fig4}\,(a) and (b) where we present typical combined and HIRO-like transitions, constructed using Eq.\,(\ref{res2}) for $(\eac,\edc)=(5/8,3/8)$ (maximum) and $(\eac,\edc)=(5/8,5/8)$ (minimum), respectively.
Here, dark and light inclined lines represent centers of real Landau levels and proposed microwave induced sidebands, respectively.
Combined transitions (dotted arrows) consist of microwave absorption (vertical arrows) and impurity scattering (horizontal arrows).
For the maximum at $(\eac,\edc)=(5/8,3/8)$ we observe that the combined transitions parallel to the dc field end at the Landau level center and therefore are maximized. 
At the same time, transitions anti-parallel to the dc field end in the gap of the density of states and are minimized.
For the minimum at $(\eac,\edc)=(5/8,5/8)$ transitions parallel to the dc field end in the middle of the gap but the antiparallel transitions terminate at the same Landau level and thus should be disregarded because emission and absorption processes compensate each other.
Contribution from impurity scattering processes which are not accompanied by microwave absorption can be safely ignored since these are neither minimized nor maximized.

Another possible scenario for frequency doubling is the interplay between ``combined'' single-photon processes and processes not-accompanied by microwave absorption (or two-photon processes) in the regime of separated Landau levels. 
As discussed in Ref. 19, the differential resistivity oscillations near half-integer values of $\eac=5/2, 7/2$ are strongly suppressed.
The suppression occurs because combined processes, maximized (minimized) at half-integer (integer) $\edc$, are compensated by HIRO-like processes which are maximized (minimized) at integer (half-integer) $\edc$. 
In the overlapped Landau level regime both contributions exhibit $\pm\cos(2\pi\edc)$-dependence which explains observed suppression but not frequency doubling.
However, at $\eac=1/2$ the Landau levels get separated and the respective contributions can no longer be described by simple $\cos$-like functions, but instead will be sharply peaked at $\edc=m+1/2$ and $\edc=m$, respectively, leading to the observed behavior.

In summary, we have studied nonlinear microwave photoresistivity of a high-mobility 2DES.
First, we have observed that in the vicinity of the the second cyclotron resonance subharmonic, $\eac=1/2$, the peaks in differential resistivity are well described by $\edc=m/2$ which is in vast contrast to the standard $\edc=m$ relation observed in the regime of integer MIRO, near $\eac=n$, or in the absence of microwave radiation.
Second, the overall evolution of the differential resistivity near $\eac=1/2$ closely replicates the behavior near $\eac=1$ once both $\eac$ and $\edc$ are scaled down by a factor of two.
Such similarity suggests conditions for the maxima and minima which correspond to single-photon combined transitions within the framework of the ``displacement'' model.
We speculate that these results might manifest the recently proposed microwave-induced spectral reconstruction \citep{dmitriev:2007b} or the competition between transitions involving different number of microwave photons in the regime of separated Landau levels.

We gratefully acknowledge discussions with I. Dmitriev and X. L. Lei \citep{lei:fmiro}.
This work was supported by NSF Grant No. DMR-0548014.

\end{document}